# Theoretical search for half-Heusler topological insulators

**Shi-Yuan Lin, Xiao-Bao Yang, and Yu-Jun Zhao***

Department of Physics and State Key Laboratory of Luminescent Materials and Devices,

South China University of Technology, Guangzhou 510640, China.

**Shu-Chun Wu, and Claudia Felser**

Max Planck Institute for Chemical Physics of Solids, 01187 Dresden, Germany

**Binghai Yan***

Max Planck Institute for Chemical Physics of Solids, 01187 Dresden, Germany

Max Planck Institute for the Physics of Complex Systems, 01187 Dresden, Germany

*Corresponding authors: Y.-J. Zhao (zhaoyj@scut.edu.cn) and Binghai Yan (yan@cpfs.mpg.de)

# ABSTRACT

We have performed *ab-initio* band structure calculations on more than two thousand half-Heusler compounds in order to search for new candidates for topological insulators. Herein, LiAuS and NaAuS are found to be the strongest topological insulators with the bulk band gap of 0.20 and 0.19 eV, respectively, different from the zero band gap feature reported in other Heusler topological insulators. Due to the inversion asymmetry of the Heusler structure, their topological surface states on the top and bottom surfaces exhibit *p*-type and *n*-type carriers, respectively. Thus, these materials may serve as an ideal platform for the realization of topological magneto-electric effects as polar topological insulators. Moreover, these topological surface states exhibit the right-hand spin-texture in the upper Dirac cone, which distinguish them from currently known topological insulator materials. Their topological nontrivial character remains robust against in-plane strains, which makes them suitable for epitaxial growth of films.

I. INTRODUCTION

Topological insulators (TIs) that are characterized by metallic surface states inside the bulk band gap have attracted great attention in the recent years [1-17]. Recently, various searches for new topologically nontrivial phases have been extended to ternary compounds [18-24]. In particular, many half-Heusler compounds have been predicted to be TI by band structure calculations [18, 19, 25]. Their band structures are characterized by a band inversion at the Γ point, which is similar to that of another known TI, HgTe. However, most of the reported Heusler TIs exhibit zero band gap [25-27], which may result in carriers in the bulk by thermal excitation at finite temperatures. External strains or structure distortions are required to open the bulk band gap, which is usually in the order of meV, to realize a real TI. Though a few negative spin-orbit splitting induced half-Heusler TIs (including LiAuS and NaAuS) with large gap were reported by Zunger et al [28], it is not clear whether more TIs with large gaps exist in half-Heusler compounds. Therefore, a systematic survey of new half-Heusler topological insulators with considerable band gap is strongly demanded, as well as the electronic property study of the attractive candidates.

Half-Heusler compounds (chemical formula MM'X) are usually non-magnetic and semiconducting when the number of total valence electrons is 18,

$$V_M + V_{M'} + V_X = 18,$$

*so called* 18-electron-rule. Therefore, half-Heusler compounds that satisfy this rule are natural candidates to search for TIs. Here, M and M' represent elements from group IA to group IIB except H, Cs, La series, and those in period 7. A total of 38 chemical elements are taken into account. X is chosen from group IIIA to group VIIA except F, At, Po and those in period 7 with a number of 22 elements in total. Totally we have 2295 compounds as possible candidates for TIs. We define the band inversion strength, $\Delta$, by the energy difference between the $\Gamma_6$ states and the conduction bands minimum (CBM). So negative values of $\Delta$ correspond to the topologically nontrivial materials, while the positive values represent the topologically trivial materials. Among all these materials, we found that LiAuS and NaAuS are the most interesting TIs with the bulk band gap of 0.20 and 0.19eV, respectively. In particular, they distinguish from currently known topological insulator materials with their topological surface states exhibit the right-hand spin-texture in the upper Dirac cone.

## II. COMPUTATIONS

The calculations were carried out with the spin polarized density functional theory (DFT) as implemented in Vienna *ab* initio simulation package (VASP) [29]. The Perdew-Burke-Ernzerhof type generalized gradient approximation (GGA) [30] was adopted for the exchange and correlation functionals. A plane wave cutoff of 450 eV and a 5×5×5 Monkhorst–Pack k-point mesh are used for the calculation. To confirm the reliability of our calculation, we also calculated the band inversion strength of YAuPb, of which the calculated $\Delta$ is -0.72 eV, in excellent agreement with refs. 18 and 19. All the calculations are performed with spin-orbit coupling (SOC). The crystal structure of half-Heusler compounds is described by the space group $F\bar{4}3m$, with $C1_b$ structure and the atomic arrangement is presented in Fig. 1. These compounds are assigned the chemical formula MM'X and the positions of the nonequivalent atoms in Wyckoff coordinates are: M atoms at (1/2 1/2 1/2), M' at (1/4 1/4 1/4) and X at (0 0 0) [31].

## III. RESULTS AND DISCUSSION

The calculated equilibrium lattice constants, band inversion strengths, and bulk band gaps of the systems with band gaps greater than 0.08 eV are listed in Table I. It was reported that [25, 32] the computations based on the local-density approximation(LDA) or GGA often result in a smaller value of the band inversion strength comparing with the MBJLDA [33] result. Particularly, ScAuPb and YPdBi are predicted of small negative band inversion strength in LDA, but become positive when using MBJLDA potential [32]. However, we find that the error between LDA or GGA and MBJLDA is within 0.3 eV for most of the half-Heusler candidates in the literature, and all the half-Heusler topological insulators with more than 1.0 eV band inversion strengths are predicted to be topologically nontrivial phases [25, 33]. In fact, the band inversion strengths of LiAuS and NaAuS are calculated to be -1.35 and -1.48 eV in our work, much beyond the error bar. In addition, their bulk band gaps are calculated to be 0.20 and 0.19 eV respectively, though suffering the well-known underestimation of band gaps by GGA.

The calculated band inversion results of the 2295 half-Heusler candidates MM'X are shown in Fig. 2 by the pentagons, of which the five sections are corresponding to the five possible choices for the anion(X position), from period 2 to 6 based on 18-electron-rule. Both M and M' are indexed with periods 2 to 6 of group I and group II elements, followed by the transition metals of periods 4 and 6 except Cs and La series. Since Po, F and At are not taken into account, those pentagons corresponding to group VIA and VIIA anions have only four and three sections, respectively. For example, if we choose Be and Co for M and M' respectively, X is fixed with Cl, Br, and I at group VIIA except F and At. The colored sections stand for the topologically nontrivial systems with band inversions. The various colors are corresponding to the compounds with anions at various periods as shown in the color bars in Fig. 2.

In total, there are 779 candidates are found to have band inversions, mostly distributed at the ends of the each strip of pentagons, which are separated in nine large strip regions due to the 18-electron rule. Some of these systems are already reported, such as M=Lu, La, Sc, and Y, M'=Pt, Pd, Au, and Ni, and X=Sb, Bi, and Sn [19, 25], as well as the group of (Li, Na, K, Rb)-(Au)-(O, S) and (Sc, Y, La, Lu)-(Au)-C [28]. Nevertheless, most of them are new discovered topologically nontrivial systems, for instance, LiPdCl, SrPtS and BaPtS are new candidates with large bulk band gaps. In addition, according to Fig. 2, we find that:

i). Regarding the choice of M and M', most of the half-Heusler compounds with band inversion are corresponding to the relative large difference of the valence electron number between M and M'. Especially, there are two major band inversion regions where M and M' correspond to the elements of group IA or group IIA. This is reflected in Fig. 2 by the distribution of color sections at the ends of each strip of pentagons. However, there is an exception, i.e., band inversion is rare for X=group VIIA when M and M' are corresponding to group IA and IIA. For instance, in the case of M=Na, NaHg-V, NaAu-VI, and NaPt-VII, are located at the left-top corner with colored sections except NaPtCl and NaPtBr.

ii). When X is chosen from group VIA or VIIA, more chances of band inversion in the half-Heusler compounds. There is no system with band inversion at the upper side of the middle of the strips. For instance, in the case of M=Cr, CrCo-III, CrFe-IV, and CrMn-V, are located at the upper side of the middle of the strip without colored sections, while CrCr-VI and CrV-VII are colored with one section each, corresponding to the nontrivial insulators of CrCrO and CrVBr. In general, this is in line with the rule of thumb, i.e., the anions with strong electronegativity (especially for group VIA and VIIA) are easier to have topologically nontrivial phases, though there are errors for the calculation band gaps within LDA/GGA accuracy.

iii). From points i) and ii) above, we expected that nontrivial topological insulators are likely to appear in the cases where M and M' correspond to the elements of group IA or group IIA, and the anions are chosen in group VIA. Finally, it turns out that LiAuS and NaAuS are topologically nontrivial phases with the greatest band gaps of 0.20 and 0.19eV.

We count the band inversion strengths and band gaps of these alloys, as the corresponding results shown in Fig. 3. The band inversion strengths of most of these alloys are less negative than -2 eV. Although, nearly 68% of them are found to have small band gaps, and the numbers of the alloys drops quickly with the increasing band gaps. We find that LiAuS and NaAuS are excellent topological insulators with the bulk band gaps of 0.20 and 0.19eV. Consequently, the trend of bulk band gaps of MAuS as M changes from Li, Na, K, to Rb are investigated (c.f. Fig. 4), in order to get some insight of the large gap of LiAuS. From Fig. 4, we notice that the band inversion strengths getting larger while the bulk band gaps become smaller from Li to Rb. In other words, topological insulators with lighter elements are often possessing larger bulk band gaps. This is in line with the trend of band gaps in traditional semiconductors. These rules of thumb may be helpful to guide the searching for new half-Heusler topological insulators with large bulk band gaps.

The band structures of LiAuS and NaAuS, which possess the largest band gaps in all the 2295 studied cases, are plotted in Fig. 5. The size of red dots denotes the degree of Au-$s$ orbital near the $\Gamma$ point. Near the Fermi level, the degenerate bands split into two states due to SOC, forming valence bands maximum (VBM) and CBM. In these two compounds, we note that the sign of SOC splitting is negative [28], i.e. J=3/2 state($\Gamma_8$) is below J=1/2 state($\Gamma_7$). Because of the negative SOC splitting, LiAuS and NaAuS exhibit the band gap, in contrast to other half-Heusler TIs wherein SOC splitting is positive and consequently induces the degenerate VBM and CBM at the $\Gamma$ point. As shown in Fig. 5, the $\Gamma_6$ band(Au-$s$ orbital) lies below the $\Gamma_7$ band(mainly S-$p$ orbital). This exhibits $s$-$p$ band inversion once at $\Gamma$ point. The negative value of $\Delta$ indicates the topologically nontrivial feature. Thus LiAuS and NaAuS are strong TIs with large bulk band gaps of 0.20 and 0.19 eV, showing great potential in spintronic applications.

As seen above, LiAuS has a large band gap under $s$-$p$ band inversion. We have extracted maximally localized Wannier functions [34, 35] from our *ab initio* calculations. The wave functions are projected to Au-$s$ $d$ and S-$p$ orbitals. Then we constructed a large slab model to simulate the LiAuS(111) surface, which contains 20 unit-cell layers (around 20nm thick). The top and the bottom terminations of the slab are S and Au, respectively. As shown in Fig. 6, gapless surface states exist inside the bulk band gap in the vicinity of the $\Gamma$ point. One can observe two pairs of Dirac type of surface states: one is due to the top surface (S termination) and the other is due to the bottom surface (Au termination). Because of the lack of inversion symmetry in the LiAuS slab model, the top and bottom surface states are not degenerate in energy and exhibit different energy dispersion. The fact that only one pair of surface states on a given surface (a single Fermi surface) is a strong evidence of the topologically nontrivial feature of LiAuS. As we can see, there is an emerging electric dipole field between top and bottom surface, similar to previously reported TIs, LaBiTe$_3$ [36] and BiTeCl [37]. Therefore, one can find that the surface Dirac cone is *p*-type and *n*-type doped on the S-terminated and Au-terminated surfaces, respectively, as shown in Fig. 6. Given the topologically nontrivial, the intrinsic dipole field makes these two compounds an ideal platform for the realization of topological magneto-electric effects [38]. Moreover, we observed a right-hand spin texture in the upper Dirac cone. This is opposite to previously known TI materials such as Bi$_2$Se$_3$ [12, 13, 39]. The unique spin vortex on LiAuS type TIs is attributed to the negative sign of SOC [40], which can exhibit exotic topological phenomena when interfaced with a left-hand TI material [40,41].

It is worthwhile to further investigate the strain effect on the band inversion strengths and bulk band gaps for LiAuS and NaAuS, since strains often exist in the interfaces of devices. In the simulations, the *c*-axis is unconstrained(free to relax) for a given in-plane lattice constant *a* in the ranges of 5.33 Å-6.64 Å and 5.64 Å-6.95 Å, i.e. with strains of -11.0% to 11.0% and -10.4% to 10.4%, respectively. As shown in Fig. 7, we find that the minimum of band inversion strengths is near the equilibrium lattice and it increases as the lattice constant deviates from the equilibrium. Particularly, they change rapidly when the lattice constant increases from the equilibrium value, while it changes much slower under negative strains. It is also clear that the range ability of LiAuS is much larger than that of NaAuS. In addition, the maximum of bulk band gaps of LiAuS and NaAuS are at their equilibrium lattices and the strain leads to smaller gaps. Beyond certain point (5.59 Å for LiAuS, and and 5.77 Åfor NaAuS) as the lattice decreases, the bulk band gaps of the systems become negative, indicating a transition from topological insulators into topological metals. Nevertheless, the bulk band gaps keep positive under tensile strain up to 11.0% and 10.4% for LiAuS and NaAuS, respectively.

## IV. CONCLUSION

In summary, we have studied 2295 candidates of half-Heusler alloys to search for topological insulators with large band gaps. We find some rules of thumb, i.e., band inversions often require a large difference of the valence electron numbers between M and M'. Interestingly, LiAuS and NaAuS are excellent nontrivial topological insulators with band gaps of 0.20 and 0.19eV, respectively, holding great potentials making them suitable for spintronic applications. They are also unique with their topological surface states exhibit the right-hand spin-texture in the upper Dirac cone. The band inversion strengths and bulk band gaps of these systems are found to be robust under large in-plane strains, which make them suitable for epitaxial growth of films.

## ACKNOWLEDGMENTS

This work is supported by NSFC (Grant Nos. 11174082 and 11104080), the Fundamental Research Funds for the Central Universities (Grant No. 2013ZZ0082). The computer time at National Supercomputing Center in Shenzhen (NSCCSZ), and ScGrid of the Supercomputing Center, Computer Network Information Center of CAS, are gratefully acknowledged.

**References**

[1] X. L. Qi, and S. C. Zhang, Phys. Today **63**, No. 1, 33 (2010).

[2] J. Moore, Nature (London) **464**, 194 (2010).

[3] M. Z. Hasan, and C. L. Kane, Rev. Mod.Phys. **82**, 3045 (2010).

[4] X. L. Qi, and S. C. Zhang, Rev. Mod. Phys. **83**, 1057 (2011).

[5] C. L. Kane, and E. J. Mele, Phys. Rev. Lett. **95**, 146802 (2005).

[6] B. A.Bernevig, T. L. Hughes, and S. C. Zhang, Science **314**, 1757 (2006).

[7] L. Fu, and C. L. Kane, Phys. Rev. B **76**, 045302 (2007).

[8] D. Hsieh, D. Qian, L. Wray, Y. Xia, Y. S. Hor, R. J. Cava, and M. Z. Hasan, Nature (London) **452**, 970 (2008).

[9] D. Hsieh, Y. Xia, L. Wray, D. Qian, A. Pal, J. H. Dil, J. Osterwalder, F. Meier, G. Bihlmayer, C. L. Kane, Y. S. Hor, R. J. Cava, and M. Z. Hasan, Science **323**, 919 (2009).

[10] Y. Xia, D. Qian, D. Hsieh, L. Wray, A. Pal, H. Lin, A. Bansil, D. Grauer, Y. S. Hor, R. J. Cava, and M. Z. Hasan, Nat. Phys. **5**, 398 (2009).

[11] K. Eto, Z. Ren, A. A. Taskin, K. Segawa, and Y. Ando, Phys. Rev. B **81**, 195309 (2010).

[12] D. Hsieh, Y. Xia, D. Qian, L. Wray, J. H. Dil, F. Meier, J. Osterwalder, L. Patthey, J. G. Checkelsky, N. P. Ong, A. V. Fedorov, H. Lin, A. Bansil, D. Grauer, Y. S. Hor, R. J. Cava, and M. Z. Hasan, Nature (London) **460**, 1101 (2009).

[13] H. J. Zhang, C. X. Liu, X. L. Qi, X. Dai, Z. Fang, and S. C. Zhang, Nat. Phys. **5**, 438 (2009).

[14] D. Hsieh, Y. Xia, D. Qian, L. Wray, F. Meier, J. H. Dil, J. Osterwalder, L. Patthey, A. V.

Fedorov, H. Lin, A. Bansil, D. Grauer, Y. S. Hor, R. J. Cava, and M. Z. Hasan, Phys. Rev. Lett. **103**, 146401 (2009).

[15] M. König, S. Wiedmann, C. Brüne, A. Roth, H. Buhmann, L. W.Molenkamp, X. L. Qi, and S. C. Zhang, Science **318**, 766 (2007).

[16] Y. L. Chen, J. G. Analytis, J. H. Chu, Z. K. Liu, S. K.Mo, X. L. Qi, H. J. Zhang, D. H. Lu, X. Dai, Z. Fang, S. C. Zhang, I. R. Fisher, Z. Hussain, and Z. X. Shen, Science **325**, 178 (2009).

[17] B. Yan, and S. C. Zhang, Rep. Prog. Phys. **75**, 096501 (2012).

[18] S. Chadov, X. L. Qi, J. Kübler, G. H. Fecher, C. Felser, and S. C. Zhang, Nat. Mater. **9**, 541 (2010).

[19] H. Lin, L. A. Wray, Y. Q. Xia, S. Y. Xu, S. A.Jia, R. J. Cava, A. Bansil, and M. Z. Hasan, Nat. Mater. **9**, 546 (2010).

[20] B. H. Yan, C. X. Liu, H. J. Zhang, C. Y. Yam, X. L. Qi, T. Frauenheim, and S. C. Zhang, EPL **90**, 37002 (2010).

[21] H. Lin, R. S. Markiewicz, L. A. Wray, L. Fu, M. Z. Hasan, and A. Bansil, Phys. Rev. Lett. **105**, 036404 (2010).

[22] D. Xiao, Y. G. Yao, W. X.Feng, J. Wen, W. G. Zhu, X. Q. Chen, G. M. Stocks, and Z. Y. Zhang, Phys. Rev. Lett. **105**, 096404 (2010).

[23] C. Li, J. S. Lian, and Q. Jiang, Phys. Rev. B **83**, 235125 (2011).

[24] B. H. Yan, L. Müchler, and C. Felser, Phys. Rev. Lett. **109**, 116406 (2012).

[25] W. A. Sawai, H. Lin, R. S. Markiewicz, L. A. Wray, Y. Xia, S. Y. Xu, M. Z. Hasan, and A. Bansil, Phys. Rev. B **82**, 125208 (2010).

[26] X. M. Zhang, W. H. Wang, E. K. Liu, G. D. Liu, Z. Y. Liu, and G. H. Wu, Appl. Phys. Lett. **99** 071901 (2011).

[27] S. Ouardi, C. Shekhar, G. H. Fecher, X. Kozina, G. Stryganyuk, C. Felser, S. Ueda, and K. Kobayashi, Appl. Phys. Lett. **98** 211901 (2011).

[28] J. Vidal, X. Zhang, V. Stevanović, J. W. Luo, and A. Zunger, Phys. Rev. B **86**, 075316 (2012).

[29] G. Kresse, and J. Furthmüller, Phys. Rev. B **54** 11169 (1996).

[30] J.P. Perdew, K. Burke, and M. Ernzerhof, Phys. Rev. Lett. **77** 3865 (1996).

[31] Wyckoff, R. W. G. Crystal Structures (Krieger, 1986).

[32] W. X.Feng, D. Xiao, Y. Zhang, and Y. G. Yao, Phys. Rev. B **82**, 235121 (2010).

[33] F. Tran, and P. Blaha, Phys. Rev. Lett. **102**, 226401 (2009).

[34] N. Marzari, and D. Vanderbilt, Phys. Rev. B **56**, 12847 (1997).

[35] I. Souza, N. Marzari, and D. Vanderbilt, Phys. Rev. B **65**, 035109 (2001).

[36] B. H. Yan, H. J. Zhang, C. X. Liu, X. L. Qi, T. Frauenheim, and S. C. Zhang, Phys. Rev. B **82**, 161108(R) (2001).

[37] Y. L. Chen, M. Kanou, Z. K. Liu, H. J. Zhang, J. A. Sobota, D. Leuenberger, S. K. Mo, B. Zhou, S. L. Yang, and P. S. Kirchmann, Nat. Phys. **9**, 704 (2013).

[38] X.L. Qi, T. L. Hughes, and S.C. Zhang, Phys. Rev. B **78**,195424 (2008).

[39] C. X. Liu, X. L. Qi, H. Zhang, X. Dai, Z. Fang, and S. C.Zhang, Phys. Rev. B **82**, 045122 (2010).

[40] Q. Wang, S.C. Wu, C. Felser, B. Yan, C. X. Liu, arXiv:1404.7091 (2014).

[41] R. Takahashi and S. Murakami, Phys. Rev. Lett. **107**, 166805 (2011).

TABLE I

Calculated equilibrium lattice constant ($a_0$), band inversion strengths ($\Delta$), and bulk band gaps ($E_g$) of the half-Heusler topological insulators.

| Compound | $a_0$(Å) | $\Delta$(eV) | $E_g$(eV) |
| --- | --- | --- | --- |
| LiPdCl | 6.02 | -0.85 | 0.11 |
| LiAuO | 5.61 | -2.39 | 0.14 |
| LiAuS | 5.99 | -1.35 | 0.20 |
| NaAuO | 5.98 | -2.33 | 0.16 |
| NaAuS | 6.30 | -1.48 | 0.19 |
| KAuO | 6.45 | -1.42 | 0.13 |
| KAuS | 6.77 | -0.94 | 0.12 |
| RbAuO | 6.64 | -1.28 | 0.10 |
| RbAuS | 6.99 | -0.87 | 0.10 |
| SrPtS | 6.61 | -1.63 | 0.10 |
| BaPtS | 6.88 | -1.19 | 0.16 |

**Figure captions:**

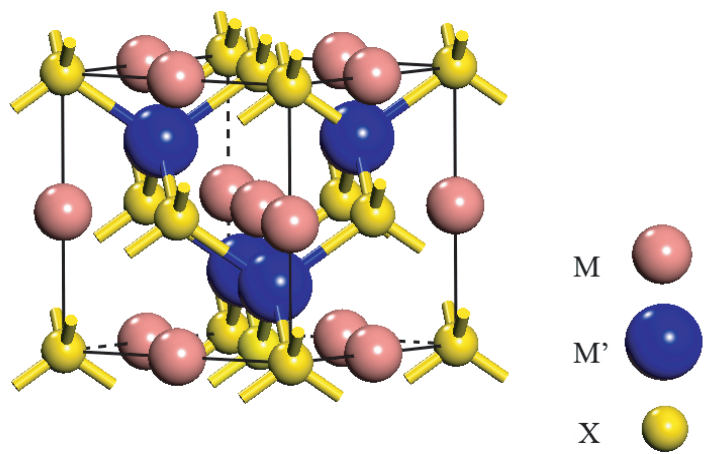

FIG. 1. The crystal structure of half-Heusler compounds MM'X.

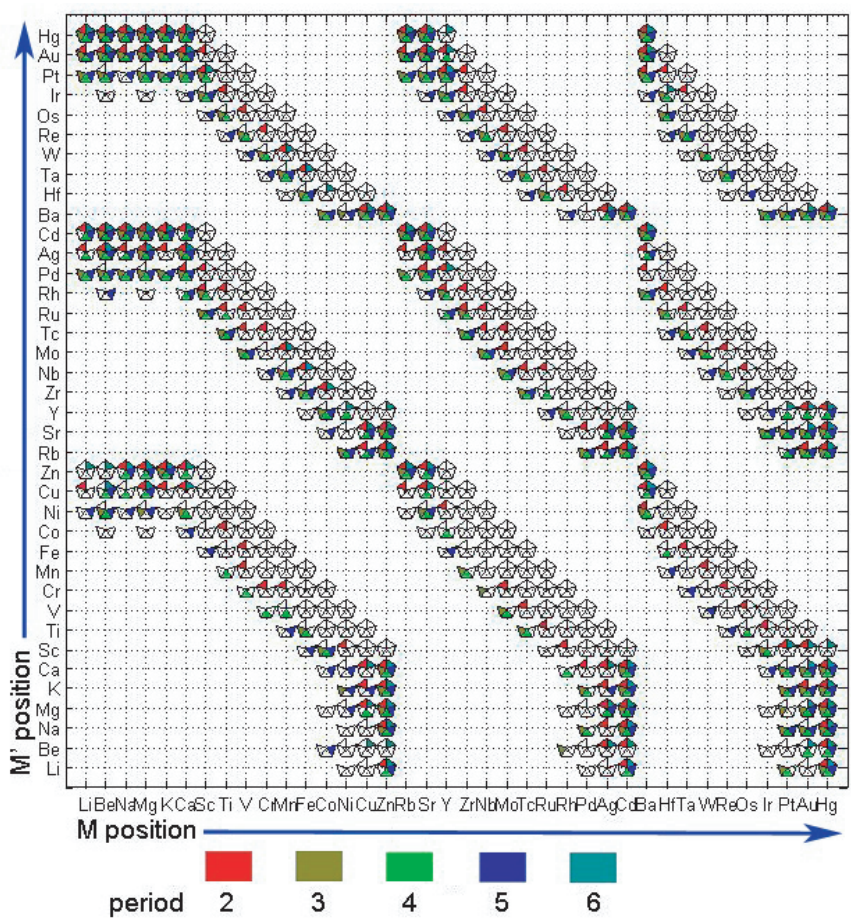

FIG. 2. The band inversion results of 2295 half-Heusler candidates MM'X. The colored

sections correspond to the systems with band inversion, i.e. TIs, as the colors stand for the anions at various periods as shown in the color bars below the figure. All the systems are shown by the pentagons in the two-dimensional graph indexed by the elements in positions M and M', when five possible choices, ranging from period 2 to 6 based on the 18-electron-rule, for anion X are represented by five sections of each pentagon. Here, both M and M' are indexed with periods 2 to 6 of group I and group II elements, followed by the transition metals of periods 4 and 6 except Cs and La series. Those pentagons with only four and three sectionsare corresponding to the compounds with anions at group VIA and group VIIA, respectively, where elements Po, F and At are not taken into account.

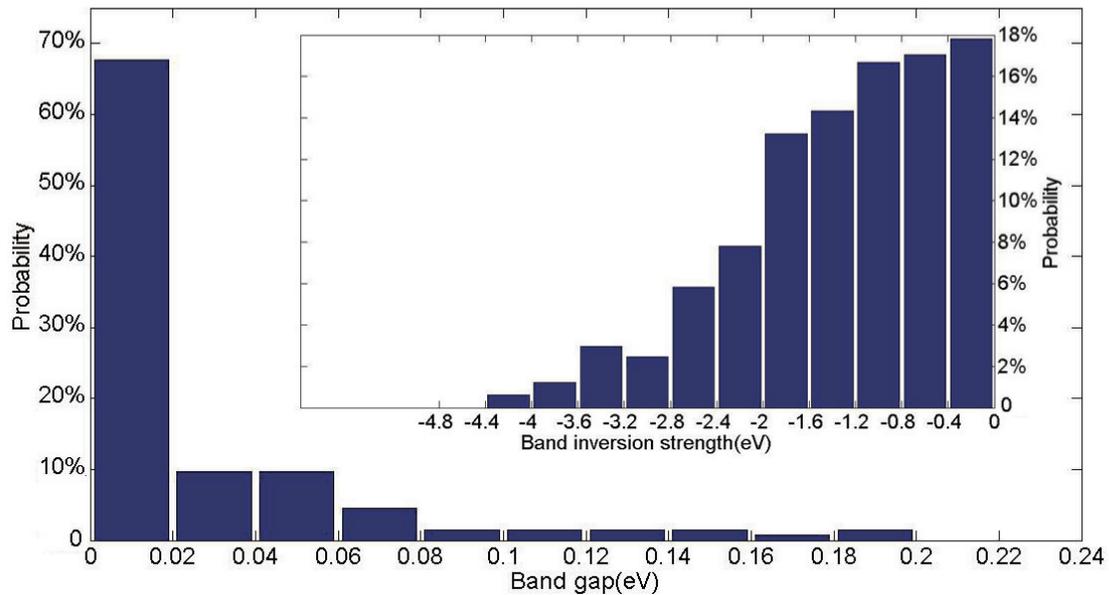

Figure 3 Lin et al

FIG. 3. The statistics of band gaps and band inversion strengths (inset panel) of the possible topologically nontrivial insulators.

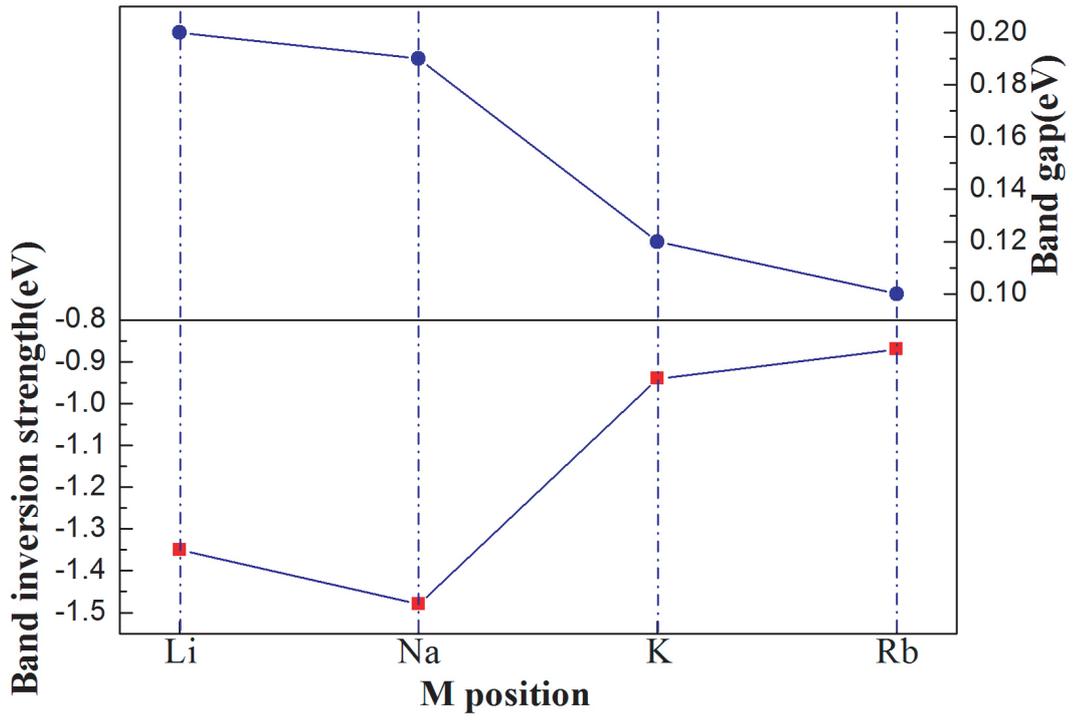

Figure 4 Lin et al

FIG. 4. The band inversion strengths and bulk band gaps of the MAuS (M=Li, Na, K, and Rb).

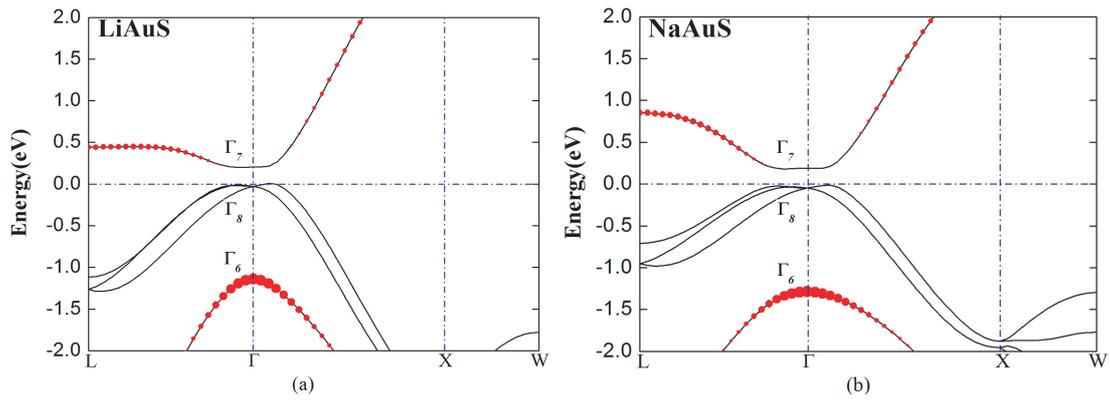

Figure 5 Lin et al

FIG. 5. The band structures of the half-Heusler topological insulators (a) LiAuS and (b) NaAuS. The horizontal dash lines indicate the VBM. Size of red dots denotes the degree of $s$-like occupancy near the $\Gamma$ point.

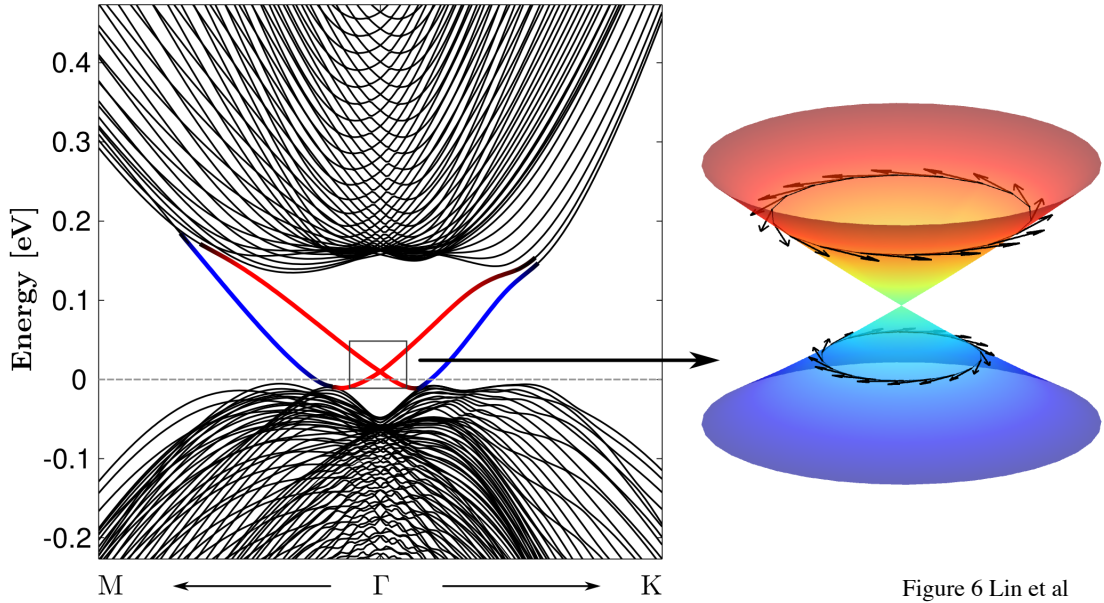

FIG. 6. Surface band structure from a slab model. The red line represents surface states from the top surface (S terminated) and the blue line represents those from bottom surface (Au terminated). The Dirac cone of the top surface is highlighted on the right panel. The right-hand spin texture exists at the upper Dirac cone.

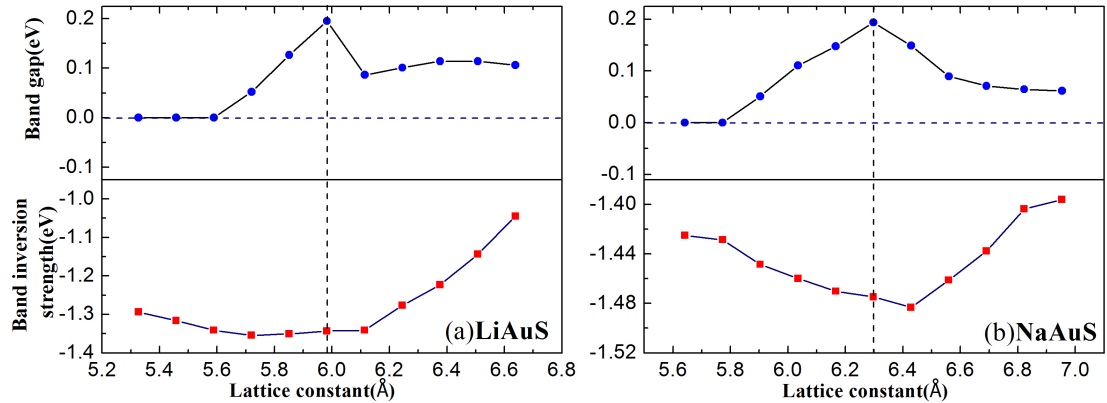

FIG. 7. The band inversion strengths and bulk band gaps as a function of the in-plane lattice constant for (a) LiAuS and (b) NaAuS. The vertical dash line indicates the equilibrium lattice constant.